\documentclass[oneside,a4paper,11pt,shownumbers]{article}
\usepackage{latexsym}
\usepackage{euscript}
\usepackage{epsfig,amsmath,amssymb}

\def\be{\begin{equation}}
\def\ee{\end{equation}}
\def\bea{\begin{eqnarray}}
\def\eea{\end{eqnarray}}

\long\def\symbolfootnote[#1]#2{\begingroup%
\def\thefootnote{\fnsymbol{footnote}}\footnote[#1]{#2}\endgroup} 

 \large\normalsize

\begin{document}

\begin{center}

{\Large \bf Charged-spinning-gravitating Q-balls}

\vspace*{7mm} {Y. Brihaye 
\symbolfootnote[1]{E-mail: yves.brihaye@umh.ac.be},
Th. Caebergs 
\symbolfootnote[1]{E-mail:thierry.caebergs@umh.ac.be}
and T. Delsate 
\symbolfootnote[1]{E-mail:terence.delsate@umh.ac.be}
}\\
\vspace*{.25cm}
{\it Facult\'e des Sciences, Universit\'e de Mons-Hainaut, 7000 Mons, Belgium}\\

\vspace*{.3cm}
\end{center}

\begin{abstract}
We consider the lagrangian of a self-interacting complex scalar field  
admitting generically Q-balls solutions. This model is extended by minimal coupling to electromagnetism and to
gravity. 
A stationnary, axially-symmetric ansatz for the different fields is used in order to reduce the 
classical equations.
The system of non-linear partial differential equations obtained becomes a boundary value
problem  by supplementing a suitable set of boundary  conditions. 
We obtain numerical evidences 
that the angular excitations of uncharged Q-balls, which exist in flat space-time, 
get continuously deformed by the Maxwell and the Einstein terms.
The electromagnetic and gravitating properties of several solutions, 
including the spinning Q-balls, are emphasized.  
\end{abstract}
\section{Introduction}
In the quest for new theoretical objects which could constitute a fraction of the exotic component (see e.g. \cite{review})
of the Universe, Q-balls \cite{fls,lp,coleman} and their gravitating counterparts, boson stars
\cite{misch,flp,jetzler}, emerge as serious candidates.
Many efforts are carried out for an experimental detection of boson stars \cite{detection}, but no evidence of their
presence has been obtained so far.
From the phenomenological point of view, the models containing the basic ingredients for Q-balls
(i.e. scalar fields) are looked for into the supersymmetric extensions of the standard model of 
particle physics \cite{dvali,kusenko,kusenko_shap,dm}. The super-partners of the quarks and leptons naturally lead to scalar 
 fields. 
 However, there is a lot of incertitude in chosing an appropriate model.
 
 Far  away from cosmological considerations and from
these elaborated lagrangians, several basic features of scalar fields in interaction
with gravity and/or electromagnetism can be studied in more elementary models. Here the basic
field is a complex scalar field self-interacting through an appropriate potential.
The spectrum of Q-balls solutions turns out to be extremely rich, the solutions 
present an harmonic time dependance characterized  by a 'spectral'-parameter $\omega_s$. 
 Besides the spherically symmetric
(fundamental) solutions, radially excitated solution exist as well\cite{vw}. The solutions can be done spinning 
both with even and  odd branches \cite{vw} under the parity operator. 
Angular excitations of Q-balls in relations with the spherical 
harmonics are discussed in \cite{bh}. Spinning solutions in supersymmetric
extensions of the standard model are constructed in \cite{cr1,cr2}.  

Once coupling the lagrangian admitting Q-balls to gravity, 
the solutions get continuously deformed by gravity but lead to an even more involved pattern \cite{kk1,kk2}. 
Several branches occur, existing on finite intervals of the parameter $\omega_s$, 
and terminating into cusps at a series of critical
values $\omega_s = \omega_{s,k}$, $k=1,2,\dots$.

Q-balls are intimately related to a global U(1) symmetry. The gauging of this symmetry naturally
introduces the electromagnetic field into the lagrangian.
It is therefore challenging to understand the pattern of solutions  to both gravitation and electromagnetism.  
The electromagnetic properties of spinning Q-balls are considered  in \cite{radu_volkov},
Charged, spherically symmetric boson stars are studied in \cite{kkll} for a very specific potential
of the scalar field. More generic potentials are proposed e.g. in \cite{st}.
To our knowledge, the effects of the electromagnetic fields on the angular excitations of boson star,
including the important case of spinning solutions have  not yet been addressed in the litterature.
The purpose of this paper is to investigate such solutions. 

In the second section of this paper, we consider the Einstein-Maxwell Lagrangian coupled to 
the Lagrangian of a scalar field admitting the basic Q-balls as generic solutions. 
We present the axially symmetric ansatz and discuss the physical quantities characterizing the solutions.

We sketch the classical equations and the set of relevant boundary conditions in Sect. III.
These equations were solved by using numerical methods, several results and properties of the
numerical solutions that we obtained are presented. The solutions can be characterized by mass,
particle number, angular momentum, electric charge and magnetic moment.
In passing, we note, that the
configurations obtained can be seen as regularized version of the Kerr-Newman black holes
where the horizon and the essential singularity occuring at the origin are regularazied by the scalar field.

\section{The model}
The starting Lagrangian decribes a complex scalar field in $3+1$ dimensions admitting Q-balls solutions.
We then couple the scalar field to an electromagnetic field in the standard way, obtaining an abelian gauge theory.
The resulting Klein-Gordon-Maxwell Lagrangian is then coupled minimally to gravity.
The full action $S$ reads:
\begin{equation}
\label{action}
 S=\int \sqrt{-g} \ d^4 x \left( \frac{R}{16\pi G} + {\cal L}_{m}\right)
\end{equation}
where $R$ is the Ricci scalar, $G$ represents Newton's constant and ${\cal L}_{m}$ denotes
the matter Lagrangian: 
\begin{equation}
\label{lag}
 {\cal L}_{m}= -\frac{1}{4} F_{\mu \nu}F^{\mu \nu}
 - D_{\mu} \Phi D^{\mu} \Phi^* - V(\Phi) \ \ , \ \ 
 F_{\mu \nu} = \partial_{\mu}A_{\nu}-\partial_{\nu}A_{\mu}
\end{equation}
Here $\Phi$ denotes the complex scalar fields and $D_{\mu} = (\partial_{\mu}- i e A_{\mu})$ 
is the covariant derivative with a coupling constant $e$. The signature of the metric is
$(-+++)$. The conventional form of the self-interacting potential  is used:
\begin{equation}
V(\Phi)= 
\kappa \vert\Phi\vert^6 - \beta \vert\Phi\vert^4 +
\lambda \vert\Phi\vert^2
\end{equation} 
where $\kappa$, $\beta$, $\lambda$, are positive constants.  
It was argued in \cite{vw} that a $\Phi^6$-potential is necessary for classical
$Q$-ball solutions to exist. 
The parameter $\lambda$ determines the mass of the  scalar field~: $(m_B)^2 = \lambda$.

With the purpose of comparing our results with those existing in the litterature \cite{kk1,kk2,bh}
in the limit $e=0$, the values 
\begin{equation}
\label{parameters}
 \kappa=1 \ \ , \ \ \beta=2 \ \ , \ \ \lambda=1.1 \ \  \ .
\end{equation}
are used in the explicit calculations.

The energy-momentum tensor reads:
\begin{equation}
\label{em}
T_{\mu\nu}= (D_{\mu} \Phi)^* (D_{\nu} \Phi) + (D_{\nu} \Phi)^* (D_{\mu} \Phi)  
+ F_{\mu \alpha}F_{\nu \beta} g^{\alpha \beta}
+g_{\mu\nu} {\cal L}_m
\end{equation}
The conserved Noether
current $j^{\mu}$, $\mu=0,1,2,3$, associated to the U(1)- symmetry is just
\begin{equation}
 j^{\mu}
 = -i \left(\Phi^* D^{\mu} \Phi - \Phi D^{\mu} \Phi^*\right)
\end{equation}
leading to a conserved charge $Q$ of the system  
\begin{equation}
 Q= - i \int j^t  \sqrt{-g} dr d \theta d \varphi 
\end{equation} 
The variation of the action \ref{action} with respect to various degrees of freedom,
namely the metric, the scalar and vector fields leads respectively to the Einstein
equations
\begin{equation}
\label{einstein}
 G_{\mu\nu}=8\pi G T_{\mu\nu} \ \ ,
\end{equation}
(with $T_{\mu\nu}$ given by (\ref{em})), the Klein-Gordon equations 
\begin{equation}
\label{KG}
 \left(\square + \frac{\partial V}{\partial \vert\Phi\vert^2} \right)\Phi=0 \ \ ,  
\end{equation}
and the Maxwell equations
\begin{equation}
\label{MA}
 \partial_{\mu}( \sqrt{-g} F^{\mu \nu}) =  \sqrt{-g} \ \frac{e}{2}( \Phi^* D^{\nu} \Phi +  D^{\nu} \Phi \Phi^* )   \ \ .
\end{equation}

\section{The Q-ball equations}
\subsection{Ansatz}
We will look for stationnary, axially-symmetric solutions  along the direction $z$.
For the metric, we use  Lewis-Papapetrou coordinate frame  \cite{kk1}:
\begin{equation}
 ds^2 = - fdt^2 + \frac{L}{f}\left(g (dr^2 + r^2 d\theta^2) + r^2 \sin^2 \theta (d\varphi + \frac{w}{r} dt)^2 \right)
\end{equation}
where the metric functions $f$, $L$, $g$ and $w$ are functions of $r$ and $\theta$
(N.B. we use the notation $L$ instead of the conventional one $l$ to avoid confusion in
 the next sections).
This ansatz is completed with the following form for the scalar and electromagnetic fields~:
\begin{equation}
\label{ansatz1}
\Phi(t,r,\theta,\varphi)=e^{i\omega_s t+ik \varphi} \phi(r,\theta) \  \ , \ \ 
{\cal A} = V(r,\theta) dt + A(r,\theta) \sin \theta d{\varphi}
\end{equation}
depending on three new functions $\phi,V,A$ while
 $\omega_s$ and  $k$ are constants. 
The periodicity of the scalar field $\Phi(\varphi)=\Phi(\varphi+2\pi)$ requires $k\in \mathbb{Z}$.

Inserting the ansatz in the equations (\ref{einstein}),(\ref{KG}),(\ref{MA}) leads to a system of seven non-linear partial
differential equations given in appendix. 
These equations depend on the spectral parameter $\omega_s$, on the integer $k$ and on the coupling constants $e,G$.
It is convenient to use dimensionless variables and functions. In this purpose, we set
\begin{equation}
\hat r = M_p r \ \ , \ \
\hat \omega_s = \frac{\omega_s}{M_p} \ \ , \ \ 
\hat \phi = \frac{\phi}{M_p} \ \ , \ \ 
\hat V = \frac{V}{M_p} \ \ , \ \ 
\hat A = \frac{A}{M_p} \ \ , 
\end{equation}
 where $M_p$ is a mass scale associated with the phenomenology of the underlying model 
 (see e.g. \cite{kasuya},\cite{cr2} for details). We then use
$\alpha \equiv 8 \pi G M_p^2$ and, for convenience, we drop the hat on the 
dimensionless quantities in the rest of the paper. The potential is fixed by (\ref{parameters}).
For $e=0$, the Maxwell equations decouple, the remaining equations are those of the (uncharged) boson star. 
For $k=0$, the equations for the fields $w$ and $A$ are solved trivially
by $w=0, A=0$.

Several limits of these equations have been studied over the recent years.
\begin{itemize}
\item The case $\alpha = e = 0$ was the object of \cite{vw}.
\item The case $e=0$ was studied  in \cite{kk1,kk2} and in \cite{bh}.
\item Charged-rotating solutions (i.e. $\alpha = 0$, $e \neq 0$, $k=1$) are constructed in \cite{radu_volkov}
\item The case  $k=0$, corresponding to
 spherically symmetric charged boson star, was adressed in \cite{kkll}. 
\end{itemize}
In \cite{bh}, it was argued that  different angular excitations
of (uncharged) Q-balls and boson-stars  exist in correspondance with the
symmetries of the spherical harmonic functions $Y_{\ell}^k(\theta,\varphi)$, with $-\ell \leq k \leq \ell$.
The fundamental solutions correspond to $\ell=k=0$. It is therefore natural to characterize
 the different families of solutions
(and their corresponding charged generalisations) by using the integers $\ell,k$.

\subsection{Boundary conditions}
We are looking for regular, stationary solutions  and presenting a  
symmetry (or an antisymmetry) under the reflexion $z \to -z$. The domain of integration
can then be limited to $r\in [0,\infty], \theta \in [0, \pi/2]$. 
It turns out that 
the boundary conditions for the function $\phi$ depend on the discrete symmetries of the fields, i.e.
 on the numbers $\ell, k$ . 
Here we give the ones corresponding to $\ell =0,1$ only (higher values can be found in \cite{bh2}). 
The appropriate boundary conditions
read:
\begin{equation}
r=0 \ \ : 
\ \partial_r f =0 \ , 
\ \  \partial_r L=0 \ , 
\ \ g=1 \ , 
\ \ w=0 \ , 
\ \  \phi=0 \ \ , 
\ \ \partial_r V = 0 \ \ ,
 \ \ A = 0 .
\label{bc3} \end{equation}
for solutions with $k\neq 0$, while for $k=0$ solutions, we have $\partial_r \phi|_{r=0}=0$.
The boundary conditions at infinity result from the requirement of asymptotic
flatness and finite energy solutions: 
\begin{equation}
r = \infty \ \ :
f =1 \ , \ \ 
L =1 \ , \ \ 
g =1 \ , \ \ 
w =0 \ , \ \ 
\phi=0 \ , \ \ 
V = 0 \ , \ \ 
A = 0 \ .
\label{bc4} \end{equation}

For $\theta=0$ the regularity of the solutions on the $z$-axis requires:
\begin{equation}
\theta = 0 \ \ :
\partial_{\theta} f=0 \ , \ \ 
\partial_{\theta} L=0 \ , \ \ 
g|_{\theta=0}=1 \ , \ \ 
\partial_{\theta} w =0 \ , \ \ 
\phi =0 \  \ , 
\ \ \partial_{\theta} V = 0 \ \ ,
 \ \ A = 0 .
\label{bc5} \end{equation}
for $k\neq 0$ solutions, while for $k=0$ solutions, we have $\partial_{\theta} \phi|_{r=0}=0$.

The conditions at $\theta=\pi/2$
are either given by
\begin{equation}
\theta = \frac{\pi}{2} \ \ :
\partial_{\theta} f=0 \ , \ \ \
\partial_{\theta} L=0 \ , \ \ \
\partial_{\theta} g=0 \ , \ \ \
\partial_{\theta} w =0 \ , \ \ \
\partial_{\theta} \phi =0 \ \ , \ \ 
 \partial_{\theta} V = 0 \ \ ,
 \partial_{\theta} A = 0 \ \ ,
\label{bc6} \end{equation}
for even parity solutions,
while for odd parity solutions the conditions for the scalar field functions read:
$\phi |_{\theta=\pi/2}=0$.

\subsection{Physical quantities}
The boson stars 
can be characterized by several physical parameters.
The mass $M$ and total angular momentum $J$ of the solution can be computed from
the appropriate Komar integral; it was shown in \cite{kk1} that
these conserved charges can finally be read off from the asymptotic behaviour
of the metric functions :
\begin{equation}
 M=\frac{1}{2G} \lim_{r\rightarrow \infty} r^2 \partial_r f  \ \ , \ \
J=\frac{1}{2G} \lim_{r\rightarrow \infty} r^2 w   \ .
\end{equation}
The total angular momentum $J$ and the Noether charges $Q$ 
of the two boson stars are related according to $J=k Q$ as pointed out first in \cite{shu_mie}.
Boson stars with $k=0$ have thus
vanishing angular momentum. 
The goal of this paper is to emphasize the
electromagnetic properties of the solutions. The electric charge $Q_e$ and  magnetic moment $\mu$
are given by the asymptotic decay of the electromagnetic potentials \cite{radu_volkov}, i.e.
\begin{equation}
    V =    \frac{Q_e}{4 \pi r} \ \ , \ \ A = - \frac{\mu}{4 \pi r} \ \ , \ \ {\rm for } \ \ r \to \infty
\end{equation}
and the gyromagnetic factor $g$ can then be defined~: ${\mu} = g Q J / (2 M)$.

\section{Numerical results}
We have solved numerically the system of partial differential equations
(\ref{einstein}) and (\ref{KG}) subject to the appropriate boundary
conditions given in Section 3.2. This has been done using the PDE solver FIDISOL
\cite{fidi}.
We have mapped the infinite interval of the $r$ coordinate $[0:\infty]$ to the
finite
compact interval $[0:1]$ using the new coordinate $z:=r/(r+1)$.  
We have typically used grid sizes of $150$ points in $r$-direction and $70$
points in $\theta$ direction. The solutions presented below have relative errors of order $10^{-3}$
or smaller.


\subsection{Non-rotating solutions}
\begin{figure}[!htb]
\centering
\leavevmode\epsfxsize=10.0cm
\epsfbox{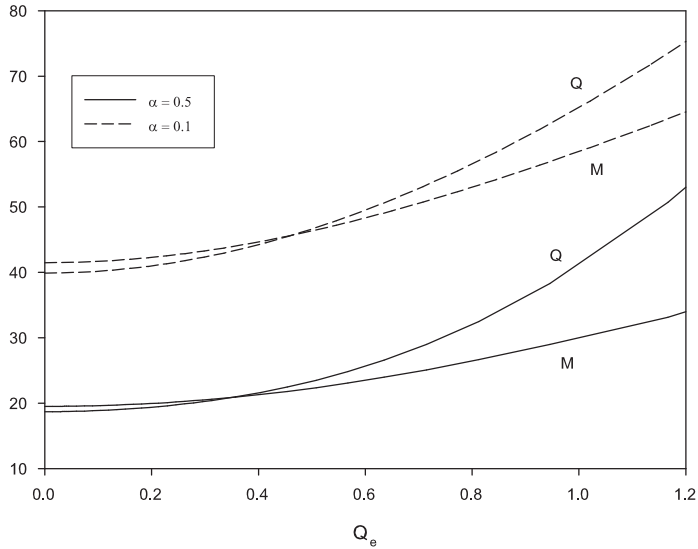}\\
\caption{\label{for_00} Values of $M$ and $Q$ as functions of the electric charge $Q_e$ for
$\alpha = 0.1$ and $\alpha = 0.5$ for the spherically symmetric boson star with $\omega_s=0.85$ }
\end{figure}

This corresponds to $k=0$. In this case, the system of differential equations
reduces to a system of coupled ordinary differential equations. Four of the equations are 
solved trivially by
$g\equiv 1$, $L=1$, $w \equiv 0$ and $A=0$, implying a null magnetic field.
The functions $f,\phi,V$ depend on the radial variable $r$ only.
In the uncharged case (i.e. with $e=0$), the gravitating Q-balls corresponding to the potential (3),(4)
 exist on a finite interval of the parameter $\omega_s$, i.e. for $\omega_s \in [\omega_1,\omega_2]$. 
 Several supplementary branches of solutions further
 develop in the region of  $\omega_s \sim \omega_a$, forming a set of backbending branches stopping
 in cusp at several critical values of $\omega_s$.
We expect this pattern of solutions to be preserved for $e > 0$. The evolution of the mass and of the
conserved charge $Q$ as functions of the electric charge $Q_e$ is reported on Fig. \ref{for_00}
for $\omega_s = 0.85$ and $\alpha=0.1, \alpha= 0.5$. For these values, there is only one uncharged solution \cite{kk1}.
Because charged spherically symmetric solutions
were first adressed in \cite{kkll} (although with a potential different from (3)), we
put in this paper the main emphasis on several kind of angular excitations, especially to the
important case of spinning solutions corrsponding  with $k = 1$.

\subsection{Spinning solutions}
\begin{figure}[!htb]
\centering
\leavevmode\epsfxsize=10.0cm
\epsfbox{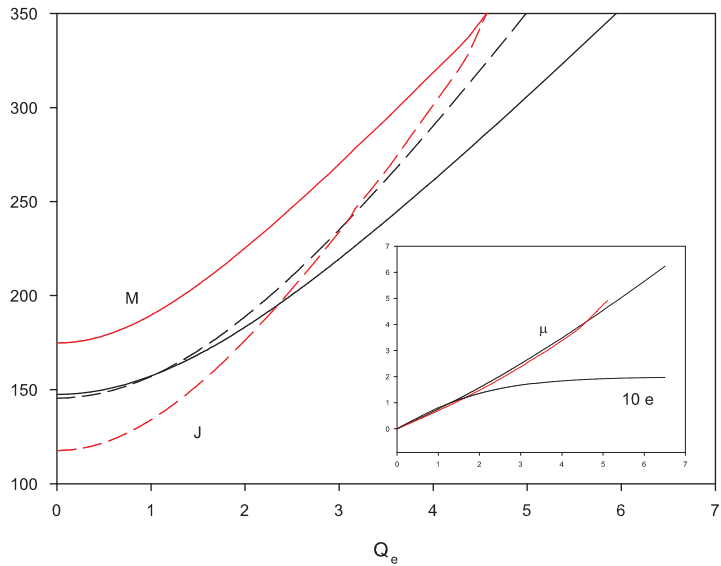}\\
\caption{\label{fig2} Mass $M$ and angular momentum $J$ of the $\ell=k=1$ Q-ball as functions of the electric charge for
$\alpha = 0.1$ and $\omega_s=0.75$ }
\end{figure}

\begin{figure}[!htb]
\centering
\epsfbox{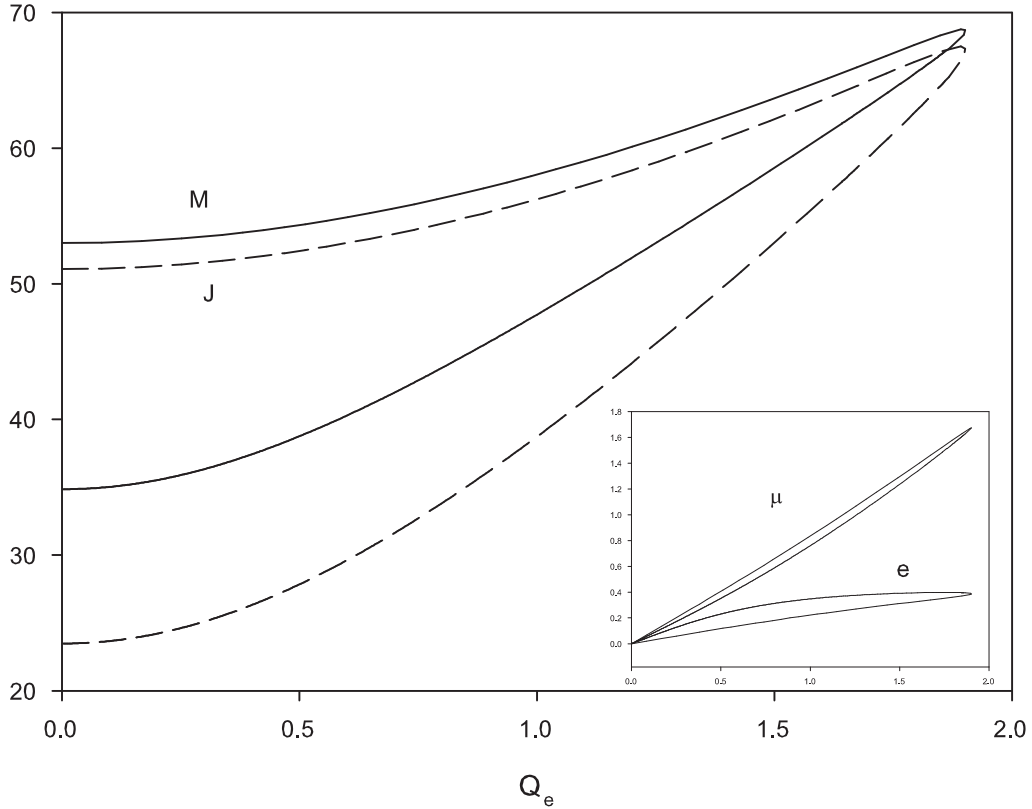}\\
\caption{\label{fig1} Mass $M$ and angular momentum $J$ of the $\ell=k=1$ Q-ball
as functions of the electric charge $Q_e$ for
$\alpha = 0.5$ and $\omega_s=0.75$ }
\end{figure}

Setting $k=1$, the solutions are charged-spinning-gravitating Q-balls.  It was shown in \cite{kk1} that 
several branches of uncharged spinning solutions exist once $\alpha > 0$. Two branches
of solutions exist on a large interval of the parameter $\omega_s$, but additionnal branches
further develop with a smaller extension of the parameter $\omega_s$.
The equations depend on three continuous parameters: $\alpha, e, \omega_s$. Exploring these
three independant directions represent a huge task which is not the scope of this paper.
We therefore limited our investigation to   $\omega=0.75$ and $\alpha \in [0.1,0.5]$
for which there are two branches of uncharged spinning boson stars \cite{kk1}.
Once increasing gradually the electric coupling constant $e$  the
 solutions naturally get charged and develop  both an electric and a magnetic fields.
The two uncharged solutions form two branches, each labelled by the electric charge $Q_e$.
Examining first the case of weak gravity, it seems that the two branches can be continued for
arbitrarily large values of the electric charge. The mass and angular momentum of the
solutions is represented on Fig. \ref{fig2} as functions of the electric charge $Q_e$. 
The magnetic momentum $\mu$ and the corresponding constant $e$ are supplemented in the window of the figure.
Considering then the spinning solutions for large values of $\alpha$ leads to a different scenario.
It turns out, indeed, that the two branches exist up to a maximal value of the electric charge,
say  $Q_e = Q_{e}^{max}$.
Once this limit is attained the two branches coincide and it is very likely that
no solution exist for $Q_e > Q_{e}^{max}$. In the case $\omega_s = 0.75, \alpha=0.5$, we find
$Q_{e}^{max} \approx 1.9$. This is illustrated by Fig. \ref{fig1}; 
we see that  the electric charge links the branches of configuration emerging from the two (uncharged)
solutions of \cite{kk1}.  As an attempt to interpret this phenomenon, we remember that, for a given value of $\omega_s$,
the mass of the soliton decreases for increasing $\alpha$.
This suggests that the mass of the soliton becomes too weak for the attractive gravitational force to
dominate the electrostatic repulsion inside the soliton.

\begin{figure}[!htb]
\centering
\leavevmode\epsfxsize=11.0cm
\epsfbox{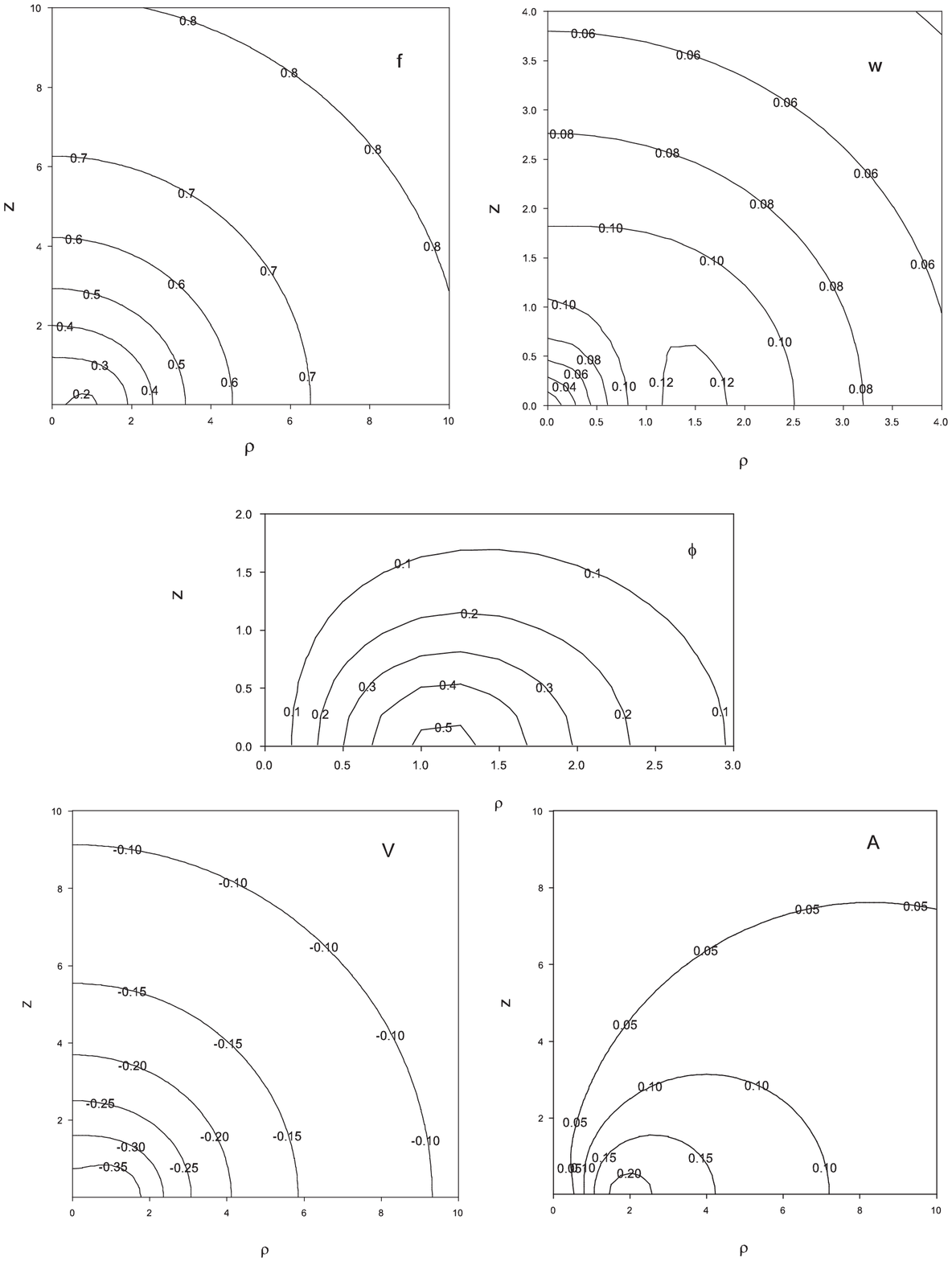}\\
\caption{\label{fig3} Contour plots of the functions $f,w,\phi, V$ and $A$ for the $\ell=k=1$
solution with
$\alpha = 0.5$ $Q_e = 1$ and $\omega_s=0.75$ }
\end{figure}

The contour plots of a
typical charged-spinning solution in the $\rho,z$ plane are presented in Fig.\ref{fig3} for $\alpha = 0.5$.
It reveals that the functions $\phi,w,A$ are concentrated in a torus lying in the equator plane, while
$f,V$ deviate only a little from spherically symmetric configurations.

\subsection{Angular excited boson stars}
Radial excitations of Q-balls have been emphasized for a long time \cite{vw},
the argument demonstrating their existence is based    on the analogy between the classical movement of a point
particle into a potential and the stationnary radial equation of the spherically-symmetric boson star  
(interpreting the radial variable $r$ as time). In \cite{bh}
it was argued that Q-balls possess angular excitations as well and that the  solutions should
exist with an angular dependance in correspondance with the spherical harmonic functions $Y_{\ell}^k(\theta,\varphi)$. 
The gravitating counterparts of these excited solutions was adressed in \cite{kk2},\cite{bh2}. 
As pointed out in the previous sections,
the cases $\ell=k=0$ and $\ell=k=1$ respectively correspond to spherically symmetric solutions and spinning solutions. 
The case $\ell=2,k=1$ corresponds to spinning solution with function $\phi$
antisymmetric under the  reflexion $z \to -z$ which we discuss in the next section. 

\begin{figure}[!htb]
\centering
\leavevmode\epsfxsize=10.0cm
\epsfbox{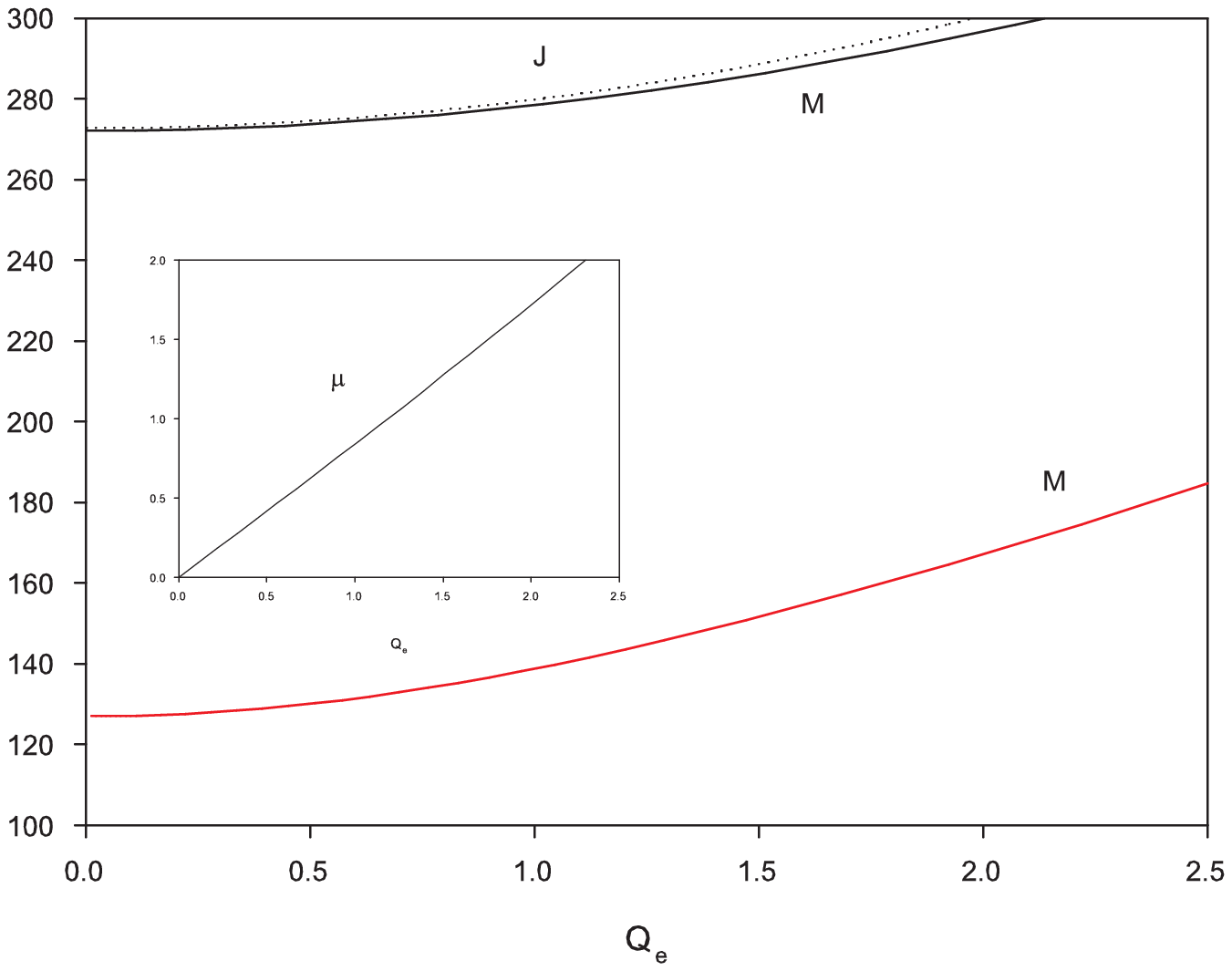}\\
\caption{\label{mass21} Mass of the $\ell=0, k=1$ solution as function of $Q_e$ (red line).
Mass, angular momentum (black lines) and magnetic momemt (in the window) of the $\ell =2, k=1$ 
solution corresponding to
$\alpha = 0.1$ and $\omega_s=0.75$ }
\end{figure}
\begin{figure}[!htb]
\centering
\leavevmode\epsfxsize=10.0cm
\epsfbox{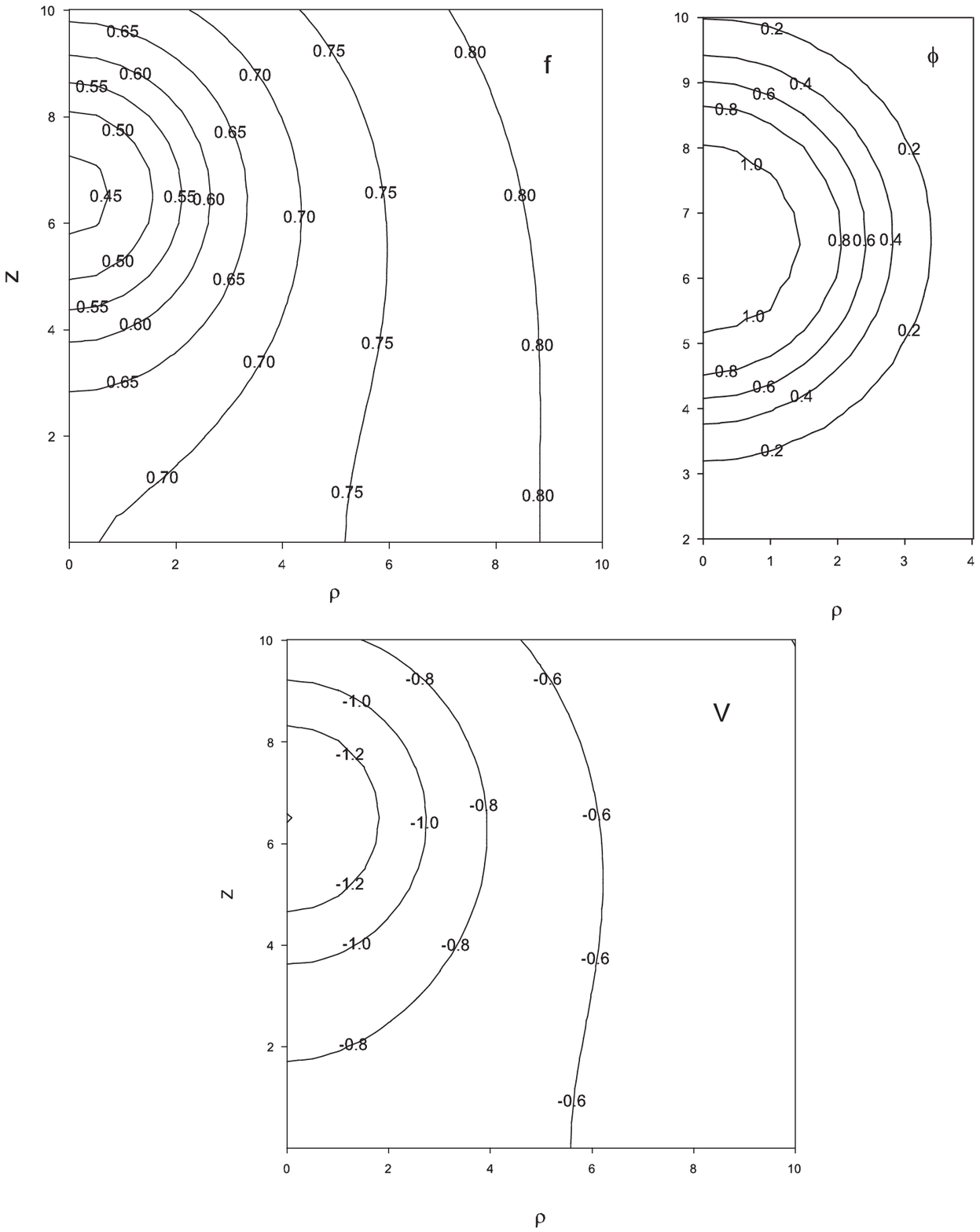}\\
\caption{\label{fig4} Contour plots of the functions $f,\phi$ and $V$ for the $\ell=1, k=0$
solution for
$\alpha = 0.5$, $\omega_s=0.75$ and $Q_e = 1$ }
\end{figure}
In \cite{bh} solutions corresponding to $\ell=1,k=0$ and $\ell=2,k=2$ have been constructed. 
Here we devote some attention to the charge and gravitating version of  the $\ell=1,k=0$ excited solution.
The angular dependance in $\theta$ goes roughly like $|Y_1^0|\propto \cos \theta$. Accordingly
it corresponds to an antisymmetric configuration of the boson field and $|\phi|$ presents  maxima 
at two opposite points along the $oz$ axis. 
The electromagnetic current is even since it is propotionnal to the product  $\phi \phi^*$.
For $k=0$, the  equations for the fields $w,A$ are trivially satisfied by $w=A=0$;  the solution
is therefore non-spinning. The mass of this solution (for $\omega_s = 0.75$, $\alpha = 0.1$) is presented as 
a function of the charge on 
Fig.\ref{mass21}. The contour plots of the solutions having $Q_e=1$ are shown on Fig. \ref{fig4}.
The maxima of the energy density are localized in two regions centered about the points $(x,y,z)=(0,0,\pm 6.5)$.
The functions $f,V$ are even, revealing  that the two maxima of gravitating matter roughly coincide
with two maxima of the density of electric charge.

\subsection{Parity-odd Charged spinning boson stars}
\begin{figure}[!htb]
\centering
\leavevmode\epsfxsize=10.0cm
\epsfbox{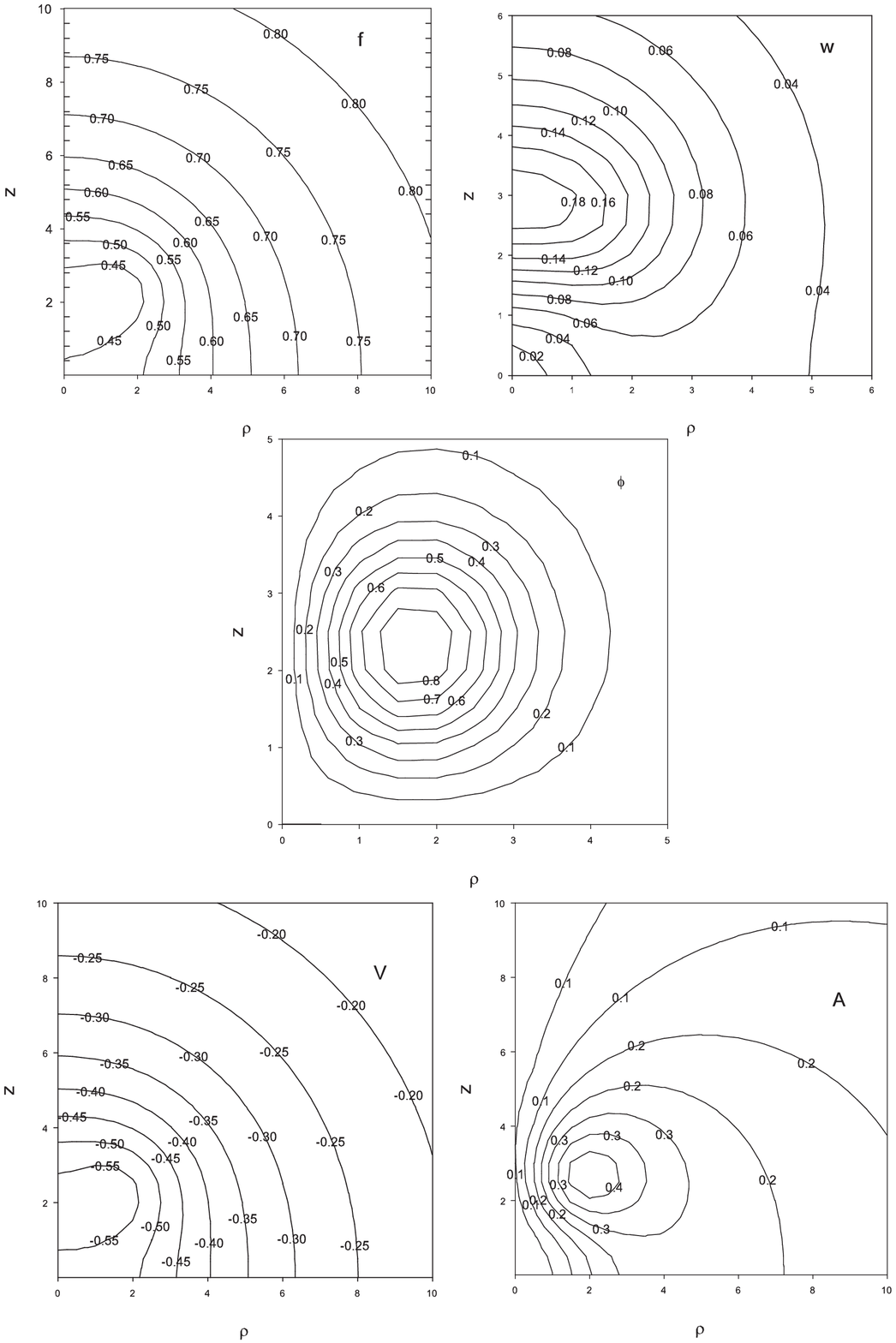}\\
\caption{\label{fig5} Contour plots of the functions $f,w,\phi, V$ and $A$ for the $\ell=2, k=1$
solution for
$\alpha = 0.1$, $\omega_s=0.75$ and $Q_e=1$ }
\end{figure}
In the section above, we discussed the charged version of fundamental parity even spinning boson stars,
it was noticed in \cite{vw} that spinning solutions with odd parity exist as well, their angular dependence
is related to the $Y_2^1$ harmonic function, with $Y_2^1 \propto \sin \theta \cos \theta$.
We succeeded in constructing the charged-gravitating version of this solution.  
The mass $M$ and angular momentum $J$ are reported on Fig.\ref{mass21} as functions of the electric charge
by the black lines. The corresponding magnetic momentum is reported in the window.
The contour plot of such a solution is shown in Fig.\ref{fig5}, we see in particular that the
scalar field is maximal on a  torus centered about the $oz$ axis and located at $z \approx  2.5$. 
Correspondingly, the field is minimal on the mirror symmetric torus at $z \approx -2.5$.
The electric and magnetic potentials are even under the reflexion $z \to -z$ 
and present extrema in the two tori. The topology of the vector potential $A$ suggests  
the structure of a double magnetic dipole.

\section{Conclusions}
Numerical arguments suggesting the existence of several families of charged gravitating,
axially symmetric Q-balls have been obtained. 
The singularities that occur naturally in the case of Einstein-Maxwell solutions
are avoided by the scalar field, so that the  solutions are regular on space-time.
We have argued that families of  charged
solitons exist for which the angular dependance of the scalar field  possesses the symmetries of the spherical harmonics
$Y_{\ell}^k(\theta,\varphi)$.
These stationnary solutions can therefore be labelled by the integers
$\ell,k$.

Asymptotically, the space-time corresponding to the 
case $\ell=k=0$ solution approaches the Reisner-Nordstrom black hole. In the case $\ell=k=1$,
the Q-balls create  a Kerr-Neumann space-time asymptotically.  
The other angular excitations generates space-times which, asymptotically, present 
 more involved angular dependances; it would be challenging
 to prove whether these solutions have (or not) black hole counterparts.  

\newpage
\newpage
\section{Appendix}
In this appendix we give the field equations in the axially symmetric ansatz.
 The notation $F^{(i,j)} \equiv \frac{\partial^i}{\partial {r^i}} \frac{\partial^j}{ \partial \theta^j} F$ is used.
  

The equations for the Maxwell fields read
\begin{equation}
 \begin{split}
&\frac{V^{(0,2)}}{r^2} + V^{(2,0)}+\frac{2 \omega  A^{(0,1)} f^{(0,1)} \sin (\theta )}{f r^3}
+\frac{2 \omega  A^{(1,0)} f^{(1,0)} \sin (\theta )}{f r}-\frac{L \omega ^2 A^{(0,1)} \omega ^{(0,1)} \sin ^3(\theta )}{f^2r^3}\\
&+\frac{L \omega ^3 A^{(1,0)} \sin ^3(\theta )}{f^2 r^2}-\frac{L \omega ^2 A^{(1,0)} \omega ^{(1,0)} \sin ^3(\theta )}{f^2 r}
-\frac{\omega  A^{(0,1)} L^{(0,1)} \sin (\theta)}{L r^3}-\frac{\omega  A^{(1,0)} L^{(1,0)} \sin (\theta )}{L r}\\
&-\frac{A^{(0,1)} \omega ^{(0,1)} \sin (\theta )}{r^3}-\frac{2 \omega  A^{(0,1)} \cos (\theta )}{r^3}-\frac{\omega^{(1,0)} \sin (\theta )}{r^2}-\frac{A^{(1,0)} \omega ^{(1,0)} \sin (\theta )}{r}\\
&+\frac{2 A \omega  f^{(0,1)} \cos (\theta )}{f r^3}-\frac{A L \omega ^2 \omega ^{(0,1)} \sin^2(\theta ) \cos (\theta )}{f^2 r^3}-\frac{A \omega  L^{(0,1)} \cos (\theta )}{L r^3}-\frac{A \omega ^{(0,1)} \cos (\theta )}{r^3}\\
&-\frac{2 e^2 L V \phi ^2 g}{f}-\frac{2 e L \phi ^2 \omega_s g}{f}-\frac{f^{(0,1)} V^{(0,1)}}{f r^2}-\frac{f^{(1,0)} V^{(1,0)}}{f}+\frac{L \omega  V^{(0,1)} \omega ^{(0,1)} \sin ^2(\theta )}{f^2 r^2}\\
&-\frac{L \omega ^2 V^{(1,0)} \sin ^2(\theta )}{f^2 r}+\frac{L \omega  V^{(1,0)} \omega ^{(1,0)} \sin ^2(\theta )}{f^2}+\frac{L^{(0,1)} V^{(0,1)}}{2 Lr^2}+\frac{L^{(1,0)} V^{(1,0)}}{2 L}\\
&+\frac{V^{(0,1)} \cot (\theta )}{r^2}+\frac{2 V^{(1,0)}}{r}-\frac{2 A \omega  \cos (\theta ) \cot (\theta )}{r^3}=0,
 \end{split}
\end{equation}


\begin{equation}
 \begin{split}
  &\frac{A^{(0,2)}}{r^2} + A^{(2,0)}+\frac{A^{(0,1)} f^{(0,1)}}{f r^2}+\frac{A^{(1,0)} f^{(1,0)}}{f}-\frac{L \omega  A^{(0,1)} \omega^{(0,1)} \sin ^2(\theta )}{f^2 r^2}+\frac{L \omega ^2 A^{(1,0)} \sin ^2(\theta ) }{f^2r}\\
&-\frac{L \omega  A^{(1,0)} \omega ^{(1,0)} \sin ^2(\theta )}{f^2}-\frac{A^{(0,1)} L^{(0,1)}}{2 L r^2}-\frac{A^{(1,0)} L^{(1,0)}}{2 L}+\frac{A^{(0,1)} \cot (\theta
   )}{r^2}\\
&-\frac{2 A e^2 L \phi ^2 g}{f}+\frac{A f^{(0,1)} \cot (\theta )}{f r^2}-\frac{A L \omega  \omega ^{(0,1)} \sin (2 \theta )}{2 f^2 r^2}-\frac{A L^{(0,1)}
   \cot (\theta )}{2 L r^2}+\frac{2 e L n \phi ^2 \csc (\theta ) g}{f}\\
&+\frac{L V^{(0,1)} \omega ^{(0,1)} \sin (\theta )}{f^2 r}+\frac{L r V^{(1,0)} \omega ^{(1,0)}
   \sin (\theta )}{f^2}-\frac{L \omega  V^{(1,0)} \sin (\theta )}{f^2}-\frac{A \csc ^2(\theta )}{r^2}=0.
 \end{split}
\end{equation}


The scalar field equation reads
\begin{equation}
 \begin{split}
&\frac{\phi^{(0,2)}}{r^2} + \phi^{(2,0)}-\frac{A^2 e^2 L \phi  \omega ^2 \sin ^2(\theta ) g}{f^2 r^2}+\frac{A^2 e^2 \phi  g}{r^2}+\frac{2 A e^2 L V \phi  \omega  \sin (\theta )g}{f^2 r}\\
&+\frac{2 A e L n \phi  \omega ^2 \sin (\theta ) g}{f^2 r^2}+\frac{2 A e L \phi  \omega  \omega_s \sin (\theta )g}{f^2 r}-\frac{2 A e n \phi  \csc (\theta ) g}{r^2}-\frac{e^2 L V^2 \phi g}{f^2}\\
&-\frac{2 e L n V \phi  \omega g}{f^2 r}-\frac{2 e L V \phi  \omega_s g}{f^2}-\frac{L n^2 \phi  \omega ^2 g}{f^2 r^2}-\frac{2 L n \phi  \omega\omega_s g}{f^2 r}-\frac{L \phi\omega_s^2 g}{f^2}+\frac{L U'(\phi ) g}{2 f}\\
&+\frac{n^2 \phi  \csc^2(\theta ) g}{r^2}-\frac{L^{(0,1)} \phi ^{(0,1)}}{2 L r^2}-\frac{L^{(1,0)} \phi ^{(1,0)}}{2 L}-\frac{\phi ^{(0,1)} \cot (\theta )}{r^2}-\frac{2 \phi^{(1,0)}}{r}=0.
 \end{split}
\end{equation}


The equations for the metric fields read
\begin{equation}
 \begin{split}
  &\frac{f^{(0,2)}}{r^2} + f^{(2,0)}-\frac{6 A f^2 \alpha  A^{(0,1)} \cot (\theta )}{L r^4}-\frac{3 f^2 \alpha  \left(A^{(0,1)}\right)^2}{L r^4}-\frac{3 f^2 \alpha  \left(A^{(1,0)}\right)^2}{L r^2}-\frac{\alpha  \omega^2 \left(A^{(0,1)}\right)^2 \sin ^2(\theta )}{r^4}\\
&-\frac{2 A \alpha  \omega ^2 A^{(0,1)} \sin (\theta ) \cos (\theta )}{r^4}+\frac{2 \alpha  \omega  A^{(0,1)} V^{(0,1)} \sin(\theta )}{r^3}-\frac{\alpha  \omega ^2 \left(A^{(1,0)}\right)^2 \sin ^2(\theta )}{r^2}+\frac{2 \alpha  \omega  A^{(1,0)} V^{(1,0)} \sin (\theta )}{r}\\
&-\frac{4 A^2 e^2 L \alpha 
   \phi ^2 \omega ^2 \sin ^2(\theta ) g}{f r^2}+\frac{8 A e^2 L V \alpha  \phi ^2 \omega  \sin (\theta ) g}{f r}+\frac{8 A e L n \alpha  \phi ^2\omega ^2 \sin (\theta ) g}{f r^2}+\frac{8 A e L \alpha  \phi ^2 \omega  \omega_s \sin (\theta ) g}{f r}\\
&+\frac{2 A \alpha  \omega 
   V^{(0,1)} \cos (\theta )}{r^3}-\frac{4 e^2 L V^2 \alpha  \phi ^2 g}{f}-\frac{8 e L n V \alpha  \phi ^2 \omega  g}{f r}-\frac{8 e L V \alpha 
   \phi ^2 \omega_s g}{f}+\frac{f^{(0,1)} L^{(0,1)}}{2 L r^2}+\frac{f^{(1,0)} L^{(1,0)}}{2 L}\\
&+\frac{f^{(0,1)} \cot (\theta
   )}{r^2}-\frac{\left(f^{(0,1)}\right)^2}{f r^2}+\frac{2 f^{(1,0)}}{r}-\frac{\left(f^{(1,0)}\right)^2}{f}-\frac{4 L n^2 \alpha  \phi ^2 \omega ^2 g}{fr^2}-\frac{8 L n \alpha  \phi ^2 \omega \omega_s g)}{f r}-\frac{4 L \alpha  \phi ^2 \omega_s^2 g}{f}\\
&-\frac{L \left(\omega^{(0,1)}\right)^2 \sin ^2(\theta )}{f r^2}+\frac{2 L \omega  \omega ^{(1,0)} \sin ^2(\theta )}{f r}-\frac{L \left(\omega ^{(1,0)}\right)^2 \sin ^2(\theta )}{f}+2 L \alpha  U(\phi) g-\frac{\alpha  \left(V^{(0,1)}\right)^2}{r^2}-\alpha  \left(V^{(1,0)}\right)^2\\
&-\frac{3 A^2 f^2 \alpha  \cot ^2(\theta )}{L r^4}-\frac{A^2 \alpha  \omega ^2 \cos ^2(\theta )}{r^4}-\frac{L \omega ^2 \sin ^2(\theta )}{f r^2}=0,
 \end{split}
\end{equation}


\begin{equation}
 \begin{split}
&\frac{L^{(0,2)}}{r^2} + L^{(2,0)}+\frac{2 L \alpha  \omega ^2 \left(A^{(0,1)}\right)^2 \sin ^2(\theta )}{f r^4}+\frac{4 A L \alpha  \omega ^2 A^{(0,1)} \sin (\theta ) \cos (\theta )}{f r^4}-\frac{4 L \alpha  \omega A^{(0,1)} V^{(0,1)} \sin (\theta )}{f r^3}\\
&+\frac{2 L \alpha  \omega ^2 \left(A^{(1,0)}\right)^2 \sin ^2(\theta )}{f r^2}-\frac{4 L \alpha  \omega  A^{(1,0)} V^{(1,0)} \sin (\theta)}{f r}-\frac{4 A f \alpha  A^{(0,1)} \cot (\theta )}{r^4}-\frac{2 f \alpha  \left(A^{(0,1)}\right)^2}{r^4}\\
&-\frac{2 f \alpha  \left(A^{(1,0)}\right)^2}{r^2}-\frac{4 A^2 e^2 L^2\alpha  \phi ^2 \omega ^2 \sin ^2(\theta ) g}{f^2 r^2}+\frac{4 A^2 e^2 L \alpha  \phi ^2 g}{r^2}+\frac{8 A e^2 L^2 V \alpha  \phi ^2 \omega \sin (\theta ) g}{f^2 r}\\
&+\frac{8 A e L^2 n \alpha  \phi ^2 \omega ^2 \sin (\theta ) g}{f^2 r^2}+\frac{8 A e L^2 \alpha  \phi ^2 \omega \omega_s \sin (\theta ) g}{f^2 r}-\frac{8 A e L n \alpha  \phi ^2 \csc (\theta ) g}{r^2}-\frac{4 A L \alpha  \omega  V^{(0,1)} \cos
   (\theta )}{f r^3}\\
&-\frac{4 e^2 L^2 V^2 \alpha  \phi ^2 g}{f^2}-\frac{8 e L^2 n V \alpha  \phi ^2 \omega  g}{f^2 r}-\frac{8 e L^2 V \alpha\phi ^2 \omega_s g}{f^2}-\frac{4 L^2 n^2 \alpha  \phi ^2 \omega ^2 g}{f^2 r^2}-\frac{8 L^2 n \alpha  \phi ^2 \omega  \omega_s g}{f^2 r}\\
&-\frac{4 L^2 \alpha  \phi ^2 \omega_s^2 g}{f^2}+\frac{4 L^2 \alpha  U(\phi ) g}{f}+\frac{2 L \alpha \left(V^{(0,1)}\right)^2}{f r^2}+\frac{2 L \alpha  \left(V^{(1,0)}\right)^2}{f}+\frac{4 L n^2 \alpha  \phi ^2 \csc ^2(\theta ) g}{r^2}\\
&+\frac{2 L^{(0,1)} \cot(\theta)}{r^2} -\frac{\left(L^{(0,1)}\right)^2}{2 L r^2} +\frac{3 L^{(1,0)}}{r}-\frac{\left(L^{(1,0)}\right)^2}{2 L}+\frac{2 A^2 L \alpha  \omega ^2 \cos ^2(\theta )}{fr^4}-\frac{2 A^2 f \alpha  \cot ^2(\theta )}{r^4}=0,
 \end{split}
\end{equation}


\begin{equation}
 \begin{split}
  &\frac{g^{(0,2)}}{r^2} + g^{(2,0)}-\frac{2 A f g \alpha  A^{(0,1)} \cot (\theta )}{L r^4}-\frac{f g \alpha  \left(A^{(0,1)}\right)^2}{L r^4}-\frac{f g \alpha  \left(A^{(1,0)}\right)^2}{Lr^2}-\frac{3 g \alpha  \omega ^2 \left(A^{(0,1)}\right)^2 \sin ^2(\theta )}{f r^4}\\
&-\frac{6 A g \alpha  \omega ^2 A^{(0,1)} \sin (\theta ) \cos (\theta )}{f
   r^4}+\frac{6 g \alpha  \omega  A^{(0,1)} V^{(0,1)} \sin (\theta )}{f r^3}-\frac{3 g \alpha  \omega ^2 \left(A^{(1,0)}\right)^2 \sin ^2(\theta )}{f r^2}\\
&+\frac{6g \alpha  \omega  A^{(1,0)} V^{(1,0)} \sin (\theta )}{f r}+\frac{6 A g \alpha  \omega  V^{(0,1)} \cos (\theta )}{f r^3}-\frac{3 g L \left(\omega^{(0,1)}\right)^2 \sin ^2(\theta )}{2 f^2 r^2}+\frac{3 g L \omega  \omega ^{(1,0)} \sin ^2(\theta )}{f^2 r}\\
&-\frac{3 g L \left(\omega ^{(1,0)}\right)^2 \sin^2(\theta )}{2 f^2}+\frac{g \left(f^{(0,1)}\right)^2}{2 f^2 r^2}+\frac{g \left(f^{(1,0)}\right)^2}{2 f^2}-\frac{3 g \alpha  \left(V^{(0,1)}\right)^2}{f
   r^2}\\
&-\frac{3 g \alpha  \left(V^{(1,0)}\right)^2}{f}-\frac{\left(g^{(0,1)}\right)^2}{gr^2}+\frac{g^{(1,0)}}{r}-\frac{\left(g^{(1,0)}\right)^2}{g}-\frac{2 g L^{(0,1)} \cot (\theta )}{L r^2}-\frac{2 g L^{(1,0)}}{Lr}\\
&-\frac{g \left(L^{(0,1)}\right)^2}{2 L^2 r^2}-\frac{g \left(L^{(1,0)}\right)^2}{2 L^2}+\frac{2 g \alpha  \left(\phi ^{(0,1)}\right)^2}{r^2}+2 g \alpha  \left(\phi ^{(1,0)}\right)^2+\frac{2 A^2 e^2 g^2 L \alpha  \phi ^2 \omega ^2 \sin ^2(\theta )}{f^2 r^2}\\
&-\frac{6 A^2 e^2 g^2 \alpha  \phi ^2}{r^2}-\frac{A^2 f g \alpha  \cot ^2(\theta )}{L r^4}-\frac{3 A^2 g \alpha  \omega ^2 \cos ^2(\theta )}{f r^4}-\frac{4 A e^2 g^2 L V \alpha  \phi ^2 \omega  \sin (\theta )}{f^2 r}-\frac{4 A e g^2 L n \alpha  \phi ^2 \omega ^2 \sin (\theta )}{f^2 r^2}\\
&-\frac{4 A e g^2 L \alpha  \phi ^2 \omega  \omega_s \sin (\theta ) }{f^2r}+\frac{12 A e g^2 n \alpha  \phi ^2 \csc (\theta )}{r^2}+\frac{2 e^2 g^2 L V^2 \alpha  \phi ^2}{f^2}+\frac{4 e g^2 L n V \alpha  \phi ^2 \omega }{f^2 r}+\frac{4 e g^2 L V \alpha  \phi ^2 \omega_s}{f^2}\\
&+\frac{2 g^2 L n^2 \alpha  \phi ^2 \omega ^2}{f^2 r^2}+\frac{4 g^2 L n \alpha  \phi ^2 \omega \omega_s}{f^2 r}+\frac{2 g^2 L \alpha  \phi ^2 \omega_s^2}{f^2}-\frac{3 g L \omega ^2 \sin ^2(\theta )}{2 f^2 r^2} -\frac{2 g^2 L \alpha   U(\phi )}{f}-\frac{6 g^2 n^2 \alpha  \phi ^2 \csc ^2(\theta )}{r^2}=0,
 \end{split}
\end{equation}


\begin{equation}
 \begin{split}
  &\frac{\omega^{(0,2)}}{r^2} + \omega^{(2,0)}-\frac{8 A f \alpha  \omega  A^{(0,1)} \cot (\theta )}{L r^4}-\frac{4 f \alpha  \omega  \left(A^{(0,1)}\right)^2}{L r^4}+\frac{4 f \alpha  A^{(0,1)} V^{(0,1)} \csc (\theta )}{Lr^3}-\frac{4 f \alpha  \omega \left(A^{(1,0)}\right)^2}{L r^2}\\
&+\frac{4 f \alpha  A^{(1,0)} V^{(1,0)} \csc (\theta )}{L r}-\frac{4 A^2 e^2 \alpha  \phi ^2 \omega  g}{r^2}+\frac{4 A e^2 V \alpha  \phi ^2 \csc (\theta ) g}{r}+\frac{8 A e n \alpha  \phi ^2 \omega  \csc (\theta ) \text{gg}(r,\theta)}{r^2}\\
&+\frac{4 A e \alpha  \phi ^2 \omega_s \csc (\theta ) g}{r}+\frac{4 A f \alpha  V^{(0,1)} \cot (\theta ) \csc (\theta )}{L r^3}-\frac{4 e n V
   \alpha  \phi ^2 \csc ^2(\theta ) g}{r}-\frac{2 f^{(0,1)} \omega ^{(0,1)}}{f r^2}\\
&+\frac{2 \omega  f^{(1,0)}}{f r}-\frac{2 f^{(1,0)} \omega ^{(1,0)}}{f}-\frac{4
   n^2 \alpha  \phi ^2 \omega  \csc ^2(\theta ) g}{r^2}-\frac{4 n \alpha  \phi ^2 \omega_s \csc ^2(\theta ) g}{r}+\frac{3 L^{(0,1)}\omega ^{(0,1)}}{2 L r^2}\\
&-\frac{3 \omega  L^{(1,0)}}{2 L r}+\frac{3 L^{(1,0)} \omega ^{(1,0)}}{2 L}+\frac{3 \omega ^{(0,1)} \cot (\theta )}{r^2}+\frac{2 \omega^{(1,0)}}{r}-\frac{4 A^2 f \alpha  \omega  \cot ^2(\theta )}{L r^4}-\frac{2 \omega }{r^2}=0,
\end{split}
\end{equation}
 

\newpage
\newpage
\begin{thebibliography}{99}
\bibitem{review} G. Jungman, M. Kamionkowski and G. Griest, Phys. Rept. {\bf 267} (1996) 195;
G. Bertone, D. Hooper and J. Silk, ibid, {\bf 405} (2005) 279.
\bibitem{fls} R. Friedberg, T. D. Lee and A. Sirlin, Phys. Rev. D {\bf 13} (1976) 2739.
\bibitem{lp} T. D. Lee and Y. Pang, Phys. Rep. {\bf 221} (1992), 251.
\bibitem{coleman}  S. R. Coleman, Nucl. Phys. B {\bf 262} (1985), 263.
\bibitem{misch} E. Mielke and F. E. Schunck, Proc. 8th Marcel Grossmann Meeting, Jerusalem, Israel, 22-27
Jun 1997, World Scientific (1999), 1607.
\bibitem{flp} R. Friedberg, T. D. Lee and Y. Pang, Phys. Rev. D {\bf 35} (1987), 3658.
\bibitem{jetzler} P. Jetzer, Phys. Rept. {\bf 220} (1992), 163.
\bibitem{detection} F. Cappella, R. Cerulli and A. Incicchitti, Eur. Phys. J. C {\bf 4} (2002) 14;
Y. Takenaga {\it et \  al.} [Super-Kamookande collaboration], Phys. Lett. B {\bf 647}  (2007) 18;
S. Cecchini {\it et \  al.} [SLIM collaboration], Eur. Phys. J. C {\bf 57} (2008) 525.
\bibitem{dvali} G. R. Dvali, A. Kusenko and M. E. Shaposhnikov, Phys. Lett. B {\bf 417} (1998) 99.
\bibitem{kusenko} A. Kusenko, Phys. Lett. B {\bf 404} (1997), 285; Phys. Lett. B {\bf 405} (1997), 108.
\bibitem{kusenko_shap} A. Kusenko and M. E. Shaposhnikov, Phys. Lett. B {\bf 418} (1998) 46.
\bibitem{dm} {\it see e.g.} A. Kusenko, hep-ph/0009089.
\bibitem{vw} M.S. Volkov and E. W\"ohnert, Phys. Rev. D {\bf 66} (2002), 085003.
\bibitem{bh} Y. Brihaye and B. Hartmann, Nonlinearity {\bf 21} (2008), 1937. 
\bibitem{cr1} L. Campanelli and M. Ruggieri, Phys. Rev. D {\bf 77} (2008), 043504. 
\bibitem{cr2} L. Campanelli and M. Ruggieri, "Spinning Supersymmetric Q-balls"' arXiv:0904.4802. 
\bibitem{kk1} B. Kleihaus, J. Kunz and M. List, Phys. Rev. D {\bf 72} (2005), 064002.
\bibitem{kk2} B. Kleihaus, J. Kunz, M. List and I. Schaffer, Phys. Rev. D {\bf 77} (2008), 064025.
\bibitem{radu_volkov} E. Radu and M. Volkov, Phys. Rept. {\bf 468} (2008) 101.
\bibitem{kkll} B. Kleihaus, J. Kunz, C. L\"ammerzahl and M. List, "Charged Boson Stars and Black Holes",
axXiv:0902.4799.
\bibitem{kasuya} S. Kasuya and F. Takahashi, Phys. Rev. D {\bf 72} (2005), 085015.
\bibitem{st} F. E. Schunck and D. F. Torres, Int. J. Mod. Phys. D {\bf 9} (2000), 601. 
\bibitem{shu_mie} F. E. Schunck and E. Mielke, in {\it Relativity and Scientific Computing}, edited by F. W. Hehl,
R. A. Putingam and H. Ruder (Spinger, Berlin, 1996) 138.
\bibitem{bh2} Y. Brihaye and B. Hartmann, Phys. Rev. D {\bf 79} (2009), 064013. 
\bibitem{fidi} W. Sch\"onauer and R. Wei\ss, J. Comput. Appl. Math. {\bf 27} (1989) 279;
M. Schauder, R. Wei\ss \ and W. Sch\"onauer, ``The CADSOL Program Package'', Universit\"at Karlsruhe, Interner
Bericht Nr. 46/92 (1992); W. Sch\"onauer and E. Schnepf, ACM
Trans. Math. Softw. {\bf 13} (1987) 333.
\end{thebibliography}
\end{document}